\documentclass[runningheads,fleqn]{svmult}
\usepackage{makeidx}   
\usepackage{graphicx}  
\usepackage{subeqnar}  
\usepackage{multicol}  
\usepackage{taphys}    
\usepackage{cite}
\usepackage{epsf}
\makeindex             
\newcommand{\capdot}{.}
\begin{document}
\title*{Quantum Phase Transitions and Collective Modes in $d$-wave Superconductors}
%
%
%
%
\toctitle{Quantum Phase Transitions and Collective Modes
\protect\newline in $d$-wave Superconductors}
%
\titlerunning{Quantum Phase Transitions in $d$-wave Superconductors}
%
\author{Matthias Vojta\inst{1}
   \and Subir Sachdev\inst{2,3}}
\authorrunning{Matthias Vojta and Subir Sachdev}
%
%
\institute{Theoretische Physik III, Elektronische Korrelationen
und Magnetismus, Institut f\"ur Physik, Universit\"at Augsburg,
D-86135 Augsburg, Germany \and Department of Physics, Yale
University, P.O. Box 208120, New Haven, CT 06520-8120, USA \and
Department of Physics, Harvard University, Cambridge, MA 02138,
USA}

\maketitle              

\begin{abstract}
Fluctuations near second-order quantum phase transitions in
$d$-wave superconductors can cause strong damping of fermionic
excitations, as observed in photoemission experiments. The damping
of the gapless {\em nodal} quasiparticles can arise naturally in
the quantum-critical region of a transition with an additional
spin-singlet, zero momentum order parameter; we argue that the
transition to a $d_{x^2-y^2} + i d_{xy}$ pairing state is the
most likely possibility in this category. On the other hand, the
gapped {\em antinodal} quasiparticles can be strongly damped by
the coupling to antiferromagnetic spin fluctuations arising from
the proximity to a Neel-ordered state. We review some aspects of
the low-energy field theories for both transitions and the
corresponding quantum-critical behavior.
In addition, we discuss the spectral properties of the collective
modes associated with the proximity to a superconductor with
$d_{x^2-y^2} + i d_{xy}$ symmetry, and implications for
experiments.
\end{abstract}


\section{Introduction}
\label{intro}

\begin{figure}[t]
\epsfxsize=4.4in
\centerline{\epsffile{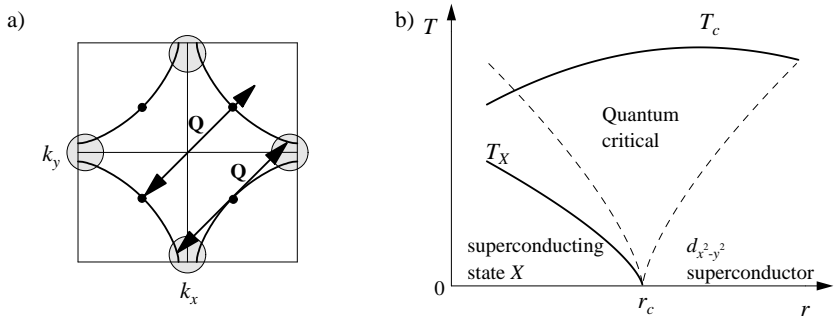}}
\caption{
a)
Location of the nodal (solid dots) and antinodal (shaded areas)
quasiparticles in the Brillouin zone.
The arrows indicate possible scattering of fermions by
order parameter fluctuations with wavevector ${\bf Q}=(\pi,\pi)$.
b)
Finite-temperature phase diagram \protect\cite{vzs}
of the $d$-wave superconductor close to a quantum critical
point.
Superconductivity is present for $T<T_c$.
The long-range order associated with the state
$X$ vanishes for $T>T_X$, but fluctuations of this order provide
anomalous damping of the nodal quasiparticles in the
quantum-critical region.
The tuning parameter $r$ is some coupling constant in the
Hamiltonian -- it is possible, although not necessary, that increasing
$r$ corresponds to increasing doping concentration, $\delta$\capdot
}
\label{fig_momentum}
\label{fig_pd}
\end{figure}

The unusual properties of quasiparticle excitations in the
cuprate high temperature superconductors, and their possible
connection with zero-temperature phase transitions in these
systems, have been the subject of intense debate over the past few
years. Angle-resolved photoemission (ARPES) experiments fail to
find well-defined quasiparticles above the superconducting
transition temperature $T_c$, and also below $T_c$ a simple
BCS-like description does not seem to apply. In this context one
has to distinguish the quasiparticles excitations in (1,1)
direction in momentum space (gapless nodal quasiparticles of the
$d$-wave superconductor) from the ones in (1,0) or (0,1)
direction (gapped antinodal quasiparticles), see
Fig.~\ref{fig_momentum}a. ARPES results~\cite{valla} indicate
that the nodal quasiparticles have very short lifetimes in the
superconducting state, with their spectral functions having
linewidths of order $k_B T$, and there is little
change~\cite{scqp,kink} in this behavior when tuning $T$ through
$T_c$. The antinodal quasiparticles are broad and ill-defined
above $T_c$ \cite{zx}, but appear to narrow significantly below
$T_c$ ~\cite{scqp,peakdiphump}, forming long-lived states with an
energy gap of 30-40 meV.

To explain these unusual properties, two main paradigms have been
suggested: either (i) the systems reflect the properties of a
fundamentally new state of matter, possibly only contiguous to the
superconducting state~\cite{z2,jan2} or (ii) the physics of the
stable ground states can be understood in the framework of
BCS-like pairing, and many of the unusual finite $T$ experimental
properties arise from nearby quantum critical points which mask
the behavior of the stable phases of the
system~\cite{CSY,castro,flux,zhang,laugh,jan}.
Here, we shall follow the second line of thought, {\em i.e.}, we assume
that the stable phases have no ``exotic'' properties or
excitations, and can (in principle) be described by an
appropriate electron Hartree-Fock/RPA/BCS theory with
perturbative corrections.

The purpose of this paper is to discuss quantum phase transitions
in the high-$T_c$ compounds, and their possible connection to the
anomalous quasiparticle properties as seen in ARPES
experiments~\cite{valla,scqp,kink,zx,peakdiphump}. For reasons
explained below, we will concentrate on transitions related to the
onset (a) of additional $i s$ or $i d_{xy}$ pairing in
$d_{x^2-y^2}$-wave superconductors (leading to damping of {\em
nodal} quasiparticles) and (b) of antiferromagnetic order
(leading to damping of {\em antinodal} quasiparticles).
Furthermore, we will discuss in more detail some properties of a
$d_{x^2-y^2} + i d_{xy}$ superconductor close to the
quantum critical point (a), and derive its collective excitation
spectrum.

The first quantum phase transition studied in the context of cuprate
superconductors was the destruction of N\'{e}el order by doping.
This transition has been proposed~\cite{CSY,science} to belong to
the same universality class as the order-disorder transition
in insulating antiferromagnets driven by quantum fluctuations~\cite{CHN}.
This implies a dynamic critical exponent $z=1$,
and a stable $S=1$ ``resonant'' spin excitation near the antiferromagnetic
wavevector in the paramagnetic state \cite{CSY,pwa},
both being consistent with numerous neutron scattering and NMR
studies.
Assuming that the mobile charge carriers have a $d$-wave
superconducting ground state on both sides of the transition,
the arguments for the common
universality of the magnetic quantum critical point in insulating
and doped antiferromagnets can be sharpened.
The order parameter is a 3-component real field, $N_{\alpha}$, which measures
the amplitude of the local antiferromagnetic order.
The onset of long-range antiferromagnetism
is described by the condensation of
$N_{\alpha}$, and the theory for the quantum critical point will
depend upon whether the $N_{\alpha}$ couple efficiently to the
low-energy fermionic excitations of the $d$-wave superconductor
(the $S=1/2$ Bogoliubov quasiparticles).
Momentum conservation now plays a key role: the dominant fluctuations
of $N_{\alpha}$ occur at the ordering wavevector ${\bf Q}$,
and the fermions will be scattered by this the wavevector.
If ${\bf Q}$ does not equal
the separation between two nodal points of the $d$-wave
superconductor
[located at momenta $(\pm K, \pm K)$ with $K = 0.39 \pi$
at optimal doping, see Fig~\ref{fig_momentum}a],
then the fermion scattering
does not lead to any disruptive low energy damping of
the $N_{\alpha}$. In such a situation, there is no
fundamental difference between the low-energy magnetic fluctuations in a
superconductor and an insulating paramagnet, and the same
theory for the antiferromagnetic quantum transition applies.
Of course, these arguments also hold if the charge sector is
fully gapped on both sides of the transition, {\em i.e.}, if the
ground state around the magnetic quantum critical point is
insulating.

The magnetic quantum phase transition alone is not likely to explain
the plethora of unusual finite-$T$ properties of the cuprates.
Instead, it appears that a variety of ordered phases compete in these
two-dimensional doped Mott insulators, and numerous additional transitions
are possible and have been discussed in recent years:
the onset of site/bond charge and/or spin density
wave order (``stripes'') \cite{castro}, additional superconducting order
like $d+is$ or $d+id$ pairing \cite{sigrist},
excitonic (or nematic) order \cite{kfe},
staggered flux (or $d$-density wave) order \cite{flux,nayak},
and ``exotic'' transitions based on the concept of spin-charge
separation \cite{z2,bfn}.
Except for the latter, all these transitions can be characterized
by an order parameter which is assumed to carry a net
momentum ${\bf Q}$ corresponding to the ordering wavevector.
Whereas superconducting and excitonic order have ${\bf Q}=0$,
stripe order is characterized by a finite (possibly incommensurate)
wavevector ${\bf Q}$, and staggered flux order corresponds to
${\bf Q}=(\pi,\pi)$ (similar to antiferromagnetic Neel order).

We now turn to the influence of these possible quantum phase transitions
on the properties of the fermionic quasiparticles,
and we will first concentrate on the nodal, {\em i.e.}, the
low-energy, fermions.
In the vicinity of a quantum transition the fermions will be strongly
scattered by bosonic order parameter fluctuations.
Similar to the discussion above, two general cases have to be
distinguished:
Either (A) the low-energy fermions are scattered into
fermion states with higher energy,
or (B) scattering occurs between low-energy fermions,
{\em i.e.}, the wavevector ${\bf Q}$ must equal the
momentum space separation between two nodal points
(nesting condition).
In case (A) fermion scattering events can be treated as virtual
processes, and the critical theory of the quantum phase transition
is not fundamentally modified by the presence of the fermions.
A perturbative expansion for the fermionic self-energy
is well behaved, and the fermion damping will vanish with a
super-linear power of temperature ($T$) as
$T \rightarrow 0$~\cite{vzs}.
In contrast, in case (B) the fermions become part of the critical
theory, a perturbative expansion in the coupling is infrared
singular, and a correct treatment requires a coupled critical theory
of bosons and fermions.
The most efficient quasiparticle scattering is provided
by a linear, non-derivative coupling between fermion bilinears
and order parameter bosons~\cite{vzs}.
If such a coupling is relevant in the renormalization group sense,
then one expects quantum-critical damping of the fermions,
{\em i.e.}, the damping rate will vanish linearly with $T$.

In Sec~\ref{sec:nodal} we review theories for
the quantum-critical damping of nodal fermions, belonging
to case (B).
We shall find that a transition to a state with
$d_{x^2-y^2} + i d_{xy}$ pairing is the most likely candidate
in this category.
The damping of antinodal fermions is subject of Sec~\ref{sec:antinodal};
we will restrict the discussion to effects arising from the proximity
to a magnetic phase transition.
Finally, in Sec~\ref{sec:clap} we return to a $d_{x^2-y^2} + i d_{xy}$
superconductor and discuss collective modes arising from fluctuations
of the additional order parameter component close to the critical
point.


\section{Damping of Nodal Quasiparticles}
\label{sec:nodal}

Motivated by the ARPES experiments on
${\rm Bi}_2 {\rm Sr}_2 {\rm Ca Cu}_2 {\rm O}_{8+\delta}$ \cite{valla}
which indicate a nodal fermion scattering rate being
proportional to $T$,
we are interested quantum-critical damping of the nodal
fermions [case (B) above].
It arises from fluctuations near a quantum critical point between the
$d_{x^2-y^2}$ superconductor and some other superconducting state
$X$ (see Fig~\ref{fig_pd}b).
The corresponding critical theories
allow for a well-controlled treatment using
renormalization group (RG) techniques.
The reason is the restricted phase space for low-energy
excitations in a $d$-wave superconductor;
in a metallic system [where case (B) is the generic situation]
the situation is much more complicated due to the presence
of low-energy particle-hole excitations at arbitrary momenta.

As discussed above, quantum-critical scattering requires
that the momentum ${\bf Q}$ carried by the order parameter
connects two nodal points.
Three natural possibilities can satisfy this
nesting condition: ${\bf Q}$ =0, ${\bf Q} =
(2K,2K)$, and ${\bf Q}=(2K, 0),(0,2K)$.
Importantly, $K$ depends on microscopic parameters,
therefore the latter two require fine-tuning, unless
there is a mode-locking between the values of ${\bf Q}$ and $K$.
Candidates for ${\bf Q} = (2K,2K)$ are Neel order and
staggered-flux order~\cite{flux,nayak}, both have ${\bf Q} = (\pi,\pi)$,
and nesting is apparently not satisfied near optimal doping.
In addition, staggered-flux order leads to a derivative coupling
between nodal fermions and the order parameter which can be shown to
be irrelevant \cite{vzs2}.
Transitions involving the onset of spin~\cite{bfn} or site/bond charge
density waves \cite{vzs} (stripes) can possibly satisfy
${\bf Q}=(2K, 0),(0,2K)$, however, this restriction on ${\bf Q}$
is not realized by the ${\bf Q}$ values observed so far.

We are left with ${\bf Q}=0$ being the only possibility which can
naturally satisfy the momentum conservation constraints for a range
of parameter values.
Assuming that the order parameter is a ${\bf Q}=0$
spin-singlet fermion bilinear (spin triplet condensation at ${\bf
Q}=0$ would imply ferromagnetic correlations which are unlikely to be
present), group theoretic arguments \cite{vzs2} permit a complete
classification of such order parameters.
The order parameter for $X$ must be built out of
the following correlators ($c_{{\bf q} a}$ annihilates an electron
with momentum ${\bf q}$ and spin $a=\uparrow , \downarrow$)
\begin{eqnarray}
\langle c_{{\bf q} a}^{\dagger}
c_{{\bf q} a} \rangle = A_{\bf q}  ~~~~\mbox{and}~~~~
 \langle c_{{\bf q} \uparrow} c_{-{\bf q}
\downarrow} \rangle = \left[ \Delta_0 (\cos q_x - \cos q_y) +
B_{\bf q} \right] e^{i \varphi},
\label{orders}
\end{eqnarray}
where $\Delta_0$ is the background $d_{x^2-y^2}$ pairing which is
assumed to be non-zero on both sides of the transition,
$\varphi$ is the overall phase of the superconducting order,
and $A_{\bf q}$ and $B_{\bf q}$ contain the possible order
parameters for the state $X$ corresponding to condensation in the
particle-hole (or
excitonic) channel or additional particle-particle pairing,
respectively.
The functions $A_{\bf q}$ and $B_{\bf q}$ can now be expanded
in terms of the basis functions of the irreducible
representation of the tetragonal point group $C_{4v}$,
and this leads to seven distinct order parameters for the
state $X$~\cite{vzs2}.
The corresponding field theories have been analyzed
recently~\cite{vzs,vzs2} by means of RG techniques,
and the results are simple and remarkable.
Only for two cases, namely the transitions
between a $d_{x^2-y^2}$-wave superconductor and
a state with either $d_{x^2-y^2}+is$ or $d_{x^2-y^2}+id_{xy}$ pairing,
there exists a fixed point describing a second-order quantum
phase transition where the fermions are part of the
critical theory.
For all other cases, we either found runaway flows of the
couplings, with no non-trivial fixed points, or a fixed
point where the fermions are decoupled from the critical degrees
of freedom.

\begin{figure}[t]
\epsfxsize=4.1in
\centerline{\epsffile{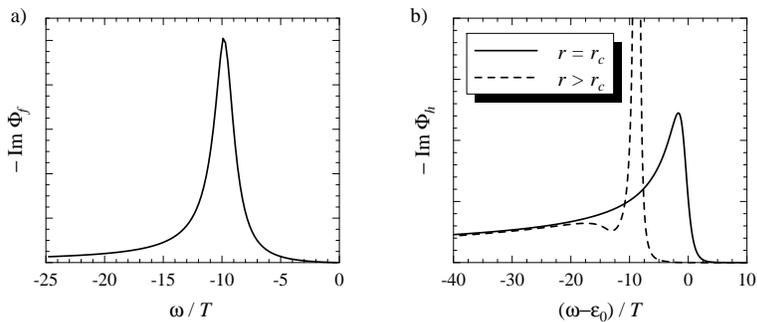}}
\caption{
Scaling functions describing the fermion damping.
a):  Fermion spectral density near a nodal point in the
quantum critical region described in Sec~\ref{sec:nodal},
in the regime $\omega,k \gg T$ (here $k/T=10$)~\cite{vzs}.
Right: Antinodal fermion spectral density near
the antiferromagnetic ordering transition (Sec~\ref{sec:antinodal}),
both in the critical region $r=r_c$ and in the regime with
gapped fluctuations $r>r_c$ ($\Delta/T=5$)~\cite{soc}.
In both cases the asymmetry of the lineshapes and the
tail at higher binding energies is clearly visible\capdot
}
\label{fig_scalf}
\end{figure}

We now describe some aspects of the critical theory
for the $d_{x^2-y^2}+is$ and $d_{x^2-y^2}+id_{xy}$ cases.
In the quantum critical region (Fig~\ref{fig_pd}b)
the single fermion Green's functions being measured
in photoemission experiments will obey the scaling form
\begin{equation}
G_f ( k , \omega) = \frac{{\cal A}_f}{T^{1-\eta_f}} \,
\Phi_f \!\left( \frac{\omega}{T}, \frac{k}{T} \right),
\label{scal}
\end{equation}
where we have set $\hbar$, $k_B$ and all velocities to unity. The
scale factor ${\cal A}_f$ is non-universal, while the anomalous
exponent $\eta_f$ and the complex-valued function $\Phi_f$ are
universal.
The nodal fermions and the order parameter fluctuations
are strongly coupled, and the
anomalous dimension of the fermion field, $\eta_f/2$, leads to
a large $\omega$ tail in its energy distribution curve
(EDC)~\cite{vzs,book,orgad}, which is actually consistent with
ARPES measurement.
The scaling function $\Phi_f$ has been studied in
Ref \cite{vzs};
for $\omega,k \gg T$, (\ref{scal}) reduces to \cite{vzs,bfn}
\begin{equation}
G_f (k, \omega) =
{\cal A} C_f \frac{ \omega + k_x \tau^z + k_y
\tau^x}{\left[ k^2 - (\omega+ i0)^2 \right]^{1-\eta_f/2}},
\label{tail}
\end{equation}
where $C_f$ is a universal number.
Note that the imaginary part of this is non-zero only for $\omega
> k$, and it decays as $\omega^{-1+\eta_f}$ for large $\omega$.
A result for the scaling function in this regime is shown in
Fig~\ref{fig_scalf}a.
In the opposite limit, $\omega,k \ll T$, (\ref{scal}) has a very
different form, and the $k$ and $\omega$ dependencies are smooth.
For further discussion we refer the reader to Ref~\cite{vzs}.

Notably, the theories for both cases, $d_{x^2-y^2}+is$ and
$d_{x^2-y^2}+id_{xy}$ pairing,
lead to identical damping behavior for the nodal fermions.
However, they can be distinguished from their effect on the
antinodal quasiparticles: $is$ fluctuations will couple
to all fermions, whereas the $id_{xy}$ order parameter has
nodes in the (1,0), (0,1) directions, therefore leaving the
antinodal fermions essentially unaffected.
With reference to the ARPES result about sharp antinodal quasiparticles
for $T \ll T_c$, we can uniquely propose the
$d_{x^2-y^2}+id_{xy}$ transition as explanation for
the nodal fermion damping below $T_c$.


\section{Damping of Antinodal Quasiparticles}
\label{sec:antinodal}

In this section, we turn to the anomalous behavior
of the antinodal fermions.
ARPES experiments show a broad EDC
above $T_c$, whereas below $T_c$ a quasiparticle-like
peak appears separated from an incoherent tail, and the spectrum
has been described as ``peak-dip-hump''
structure~\cite{scqp,zx,peakdiphump}.

For $T \ll T_c$ the antinodal quasiparticles live
at a finite energy (given by the maximum value of the
superconducting gap), and possible causes for damping
are less restricted than in the case of the nodal fermions.
We will concentrate here on fermion scattering by antiferromagnetic
fluctuations~\cite{castro,cmv};
other proposals include one-dimensional~\cite{orgad} or
two-dimensional~\cite{z2}
electron fractionalization (induced by proximity to
spin liquid states) and the
coupling to superconducting phase and vortex fluctuations~\cite{phase},
these will not be discussed here.

It can be seen from Fig~\ref{fig_momentum}a that magnetic fluctuations
with wavevector ${\bf Q}=(\pi,\pi)$ can provide an efficient coupling
between antinodal quasiparticles.
We will examine their spectral function (or EDC) in the
vicinity of a quantum critical
point between a $d$-wave superconductor and a state with
co-existing superconducting and antiferromagnetic order.
Appealing to the proximity of a magnetic quantum critical point
allows us to make controlled statements in a regime with strong
coupling between the fermionic quasiparticles and the antiferromagnetic
fluctuations \cite{soc}.

As argued in the introduction, the critical theory of the
antiferromagnetic transition itself is not influenced by
the presence of fermions, because the order parameter fluctuations, $N_\alpha$,
with momentum ${\bf Q}=(\pi,\pi)$ do not scatter between low-energy
nodal fermions of the superconductor.
Therefore, in contrast to the theories of Sec~\ref{sec:nodal},
we can work here with the ``bare'' bosonic theory for the
ordering transition, which will now be coupled to a {\em single}
anti\-nodal fermion (``hole'') \cite{soc}.
Further simplifications arise from the assumption that
the fermion momentum corresponds to an extremum or van-Hove
point of the band structure.
Then, the dispersion has no linear momentum dependence, and
RG analysis shows that quadratic and higher terms in the
dispersion are irrelevant, therefor the hole
may be viewed as dispersionless (immobile).
Also, the coupling between the charge density of the
fermion and the bulk theory is irrelevant under RG.
The resulting theory has the structure of a
Bose-Kondo-like model \cite{science,vbs} of a
single quantum spin coupled locally to the bosonic
fluctuations of a nearly critical antiferromagnet.
(A similar boundary field theory has been studied~\cite{vbs} in
the context of Zn/Li impurities in cuprate superconductors.)

We now turn to the hole spectrum as measured in photoemission.
At the $T=0$ quantum critical point at $r=r_c$,
the $N_\alpha$ are gapless critical excitations,
and a naive perturbation expansion of the hole self-energy
is infrared singular.
Our recent RG analysis \cite{soc} of the scale-invariant
quantum-field theory permits a resummation of the perturbative
expansion.
In the $r \geq r_c$, $T \geq 0$ vicinity of the critical point,
the hole Green's function obeys the scaling form
\begin{equation}
G_h(\omega) = \frac{{\cal A}_h}{T^{1-\eta_h}} \Phi_h \left(
\frac{\omega-\epsilon_0}{T}, \frac{\Delta}{T} \right)
\label{gscale}
\end{equation}
where $\Delta$ is the spin gap of the host antiferromagnet,
${\cal A}_h$ is a non-universal amplitude and
$\epsilon_0$ denotes the bare fermion energy at the antinodal point.
The anomalous exponent $\eta_h$ and the scaling function
$\Phi_h$ are again universal.
In particular, at the critical point ($r=r_c$, $T=0$) there
is no quasiparticle pole, but a power-law singularity described
by $G_h = -{\cal A}_h (\epsilon_0-\omega)^{-1+\eta_h}$
(spin orthogonality catastrope).
Numerical results for $\Phi_h$ obtained with a large $N$ method~\cite{vbs}
are displayed in Fig~\ref{fig_scalf}b.
The ${\bf k}$ dependence of $G_h$ arises only from the
irrelevant band curvature terms, and their main effect is to replace
$\epsilon_0$ by the actual hole dispersion near the van-Hove
point.
We expect that the $r=r_c$ spectrum in Fig~\ref{fig_scalf}b applies
to photoemission at the antinodal points above $T_c$.
Below $T_c$, the measured antinodal spectrum
\cite{scqp} is similar to the $r>r_c$ spectrum in Fig~\ref{fig_scalf}b:
this is accounted for in our approach by the reasonable assumption
that the onset of superconductivity induces the spin-gap-like
correlations and so increases the value of the effective $r$
controlling the magnetic fluctuations. The high frequency tail of
the EDC both above and below $T_c$ should decay as
$1/\omega^{1-\eta_h}$.


\section{Collective Modes Associated with $d_{x^2-y^2} + i d_{xy}$ Pairing}
\label{sec:clap}

In this section, we focus on the $d_{x^2-y^2} + i d_{xy}$
superconductor which emerged from our discussion of nodal
quasiparticle damping in Sec~\ref{sec:nodal}. The aim is to
identify the low-energy collective modes which should exist near
the quantum phase transition to a state with pure $d_{x^2-y^2}$
pairing, and to discuss experimental signatures of these modes.
Such collective modes have also been considered recently by
Balatsky {\em et al.} \cite{bks}, but our results below differ
significantly from theirs.

To proceed, we have to derive an effective action for the
fluctuations of the $d_{xy}$ order parameter. We will demonstrate
this explicitely on the basis of a Sp($N$) mean-field
theory~\cite{vzs} for a $t$-$J$ model with additional diagonal
exchange interaction, $J_2$, although the results discussed below
are far more general. We start in a regime where the ground state
is a pure $d_{x^2-y^2}$-wave superconductor. It is described by a
saddle point of the large-$N$ theory, and superconductivity is
encoded in non-zero pairing amplitudes on the links of the square
lattice. Now we consider a perturbation which introduces pairing
along the diagonals, with an amplitude $Q_{1,1} = - Q_{1,-1}
\equiv Q_{xy}$, where $Q_{xy}$ is a space and time dependent
complex number. We factor out the overall phase of the
superconducting order, $\varphi$, as in (\ref{orders}), and so
the phase of $Q_{xy}$ measures the {\em relative} phase between
the $d_{xy}$ and $d_{x^2-y^2}$ orders. The fluctuations of
$\varphi$ couple to overall charge fluctuations and these occur
at a high plasma frequency---we can therefore neglect $\varphi$
in our considerations. These arguments encapsulate the physics
discussed by Balatsky {\em et al.} in terms of numbers of Cooper
pairs with $d_{xy}$ and $d_{x^2-y^2}$ pairing.

After integrating out the fermions in the $d$-wave saddle point,
the large-$N$ theory gives an effective action for $Q_{xy}$ of the
following form:
\begin{eqnarray}
S_Q &=& \sum_{{\bf k}, \omega_n} \left\{
I({\bf k}, \omega_n) Q_{xy} ({\bf k}, \omega_n) Q_{xy}^\ast ({\bf k}, \omega_n)
\phantom{\frac{1}{2}} \right . \nonumber \\
~&~&~~~~~~~ \left .
+ \frac{1}{2} \left[ J({\bf k}, \omega_n) Q_{xy} ({\bf k}, \omega_n) Q_{xy}
(-{\bf k},-\omega_n) + H.c \right]
\right\},
\label{e1}
\end{eqnarray}
where $I$ and $J$ are complex functions, and have expressions
given by simple one-loop fermion diagrams.
Explicitely, we find
\begin{eqnarray}
I({\bf k}, \omega_n) &=& \frac{1}{J_2}
  - \sum_{{\bf q}, \nu_n} \gamma_{xy}^2({\bf q} - \frac{\bf k}{2}) \,
    G({\bf q},\nu_n) \, G(-{\bf q}+{\bf k},-\nu_n+\omega_n)
\nonumber \\
J({\bf k}, \omega_n) &=&
   \sum_{{\bf q}, \nu_n} \gamma_{xy}^2({\bf q} - \frac{\bf k}{2}) \,
    F({\bf q},\nu_n) \, F(-{\bf q}+{\bf k},-\nu_n+\omega_n) \,.
\label{exprab}
\end{eqnarray}
Here $G$ and $F$ are the normal and anomalous Green's functions of
the $d$-wave superconductor, and
$\gamma_{xy}({\bf q}) = \cos(q_x+q_y) - \cos(q_x-q_y)$.

\begin{figure}[t]
\epsfxsize=4.6in
\centerline{\epsffile{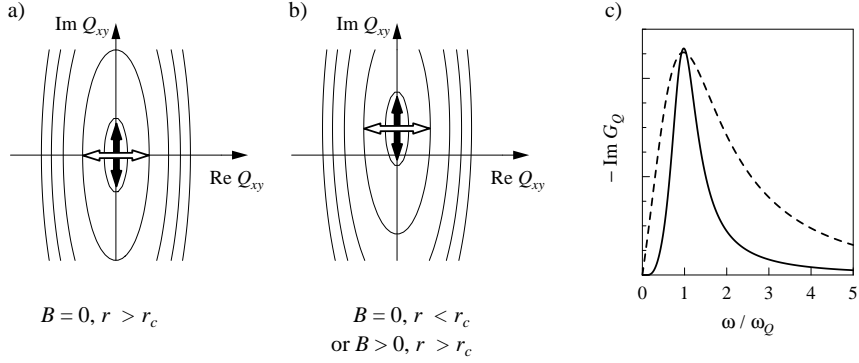}}
\caption{
a) Schematic contour plot of the Landau-Ginzburg free
energy for the fluctuations of real and imaginary part of the
$d_{xy}$ order parameter component, $Q_{xy}=\psi+i \phi$,
for zero magnetic field $B$ and $r>r_c$, {\em i.e.}, in the region where
the ground state is a pure $d_{x^2-y^2}$-wave superconductor.
The double arrows denote the normal modes of the $Q_{xy}$
fluctuations.
b) Same as a), but in the region with a $d_{x^2-y^2} + i d_{xy}$ ground state
(for $r<r_c$ or $B>0$, $r>r_c$).
c) Spectral density of the zero-momentum $Q_{xy}$ propagator, for $r>r_c$,
but close to the quantum critical point at $r_c$.
The peak corresponds to the oscillation of $\phi$ about zero,
with a characteristic energy $\omega_Q$.
Dashed: $T \ll \omega_Q$, solid: $T \gg \omega_Q$.
The excitation corresponding to the oscillation of $\psi$ is
strongly overdamped and not visible\capdot
}
\label{fig_didmod}
\end{figure}

We now study the small momentum and frequency behavior of the $Q$
fluctuations. We perform a small $k$ and $\omega_n$ expansion of
(\ref{e1}), with the decomposition $Q_{xy} = \psi + i \phi$, and
find
\begin{eqnarray}
S_Q &=& \sum_{{\bf k}, \omega_n} \left\{ \left(C_1 + C_{1k} |{\bf k}| +
C_{1 \omega} |\omega_n| \right) \psi^2 + \left(C_2 + C_{2k} |{\bf k}| +
C_{2 \omega} |\omega_n| \right)\phi^2 \right. \nonumber \\
&~&~~~~~~~~~~~~~~~~\left.+  C_3 \omega_n \psi \phi + \cdots
\right\}, \label{sq}
\end{eqnarray}
where the $C$'s are some real constants. The most notable terms
are the non-analytic $|{\bf k}|$ and $|\omega|$ terms displayed above.
The remaining higher order terms are analytic and turn into
regular gradients after Fourier transforming to the spacetime
representation.
The non-analytic terms are a consequence of the
gapless nodal quasiparticles in the $d_{x^2-y^2}$ superconductor.
Further, an analysis of the infra-red singularities of the nodal
fermion fluctuations, using the field-theoretic methods of
\cite{vzs2}, shows that the nature of the non-analyticity in
(\ref{sq}) is robust, {\em i.e.}, higher-order corrections do not
modify the non-analytic power of the momentum or frequency, or
introduce additional non-analyticities. In this sense, these
singularities are the fermionic analog of the Goldstone
singularities arising from spin waves in the ordered phase of an
antiferromagnet. Only at the quantum critical point, at which
there is onset of $d_{xy}$ order, are there additional
higher-order corrections to (\ref{sq}), which then
re-exponentiate to give the anomalous critical exponents as in
(\ref{tail}) \cite{vzs2}.

We now discuss the behavior of the low-energy action $S_Q$ near
the quantum critical point involving onset of $d_{xy}$ order. At
the transition, fluctuations described by $\phi = {\rm Im} Q_{xy}$
condense, and therefore close to the transition we expect $C_2
\ll C_1$ [note $C_{1,2} = (I\pm J)({\bf k}=0,\omega=0)$].
This situation is illustrated in Fig~\ref{fig_didmod}a,
and it is clear that the normal modes for $r>r_c$ are
oscillations of $\phi$ and $\psi$ about zero, with the $\phi$
mode having a much lower energy. However, the non-analyticities
mentioned above imply that the propagator associated with $Q_{xy}$
fluctuations does not have a simple pole structure. We have
evaluated the spectral density of this propagator in the
large-$N$ theory of Ref~\cite{vbs}, and results are shown in
Fig~\ref{fig_didmod}c. There is a peak at a characteristic
energy, $\omega_Q$, associated with the $\phi$ oscillations, but
it has a rather large width in the zero-temperature limit, $T \ll
\omega_Q$. In the opposite limit, $T \gg \omega_Q$, the peak
narrows, but does not become a simple pole. At higher energies,
the propagators are strongly damped, and no structure is visible
at an energy corresponding to $\psi$ oscillations.

Moving to $r<r_c$, ${\rm Im} Q_{xy}$ acquires a finite expectation
value, and the resulting Landau-Ginzburg free energy is
illustrated in Fig~\ref{fig_didmod}b.
Applying a non-zero magnetic field $B$ for $r>r_c$ has the same
effect (if we ignore vortex physics \cite{wolfle}), {\em i.e.},
it introduces a term proportional to $B \Delta_0 \phi$ into the action,
and thereby induces a finite $i d_{xy}$ order parameter component.
From Fig~\ref{fig_didmod}b it is clear that
the low-energy mode is still given by
amplitude fluctuations of $\phi$ whereas the $\psi$ mode is high
in energy.
Furthermore, the frequency of the $\phi$ oscillations remains $B$
independent for small fields $B$.

These results above differ from those of Ref \cite{bks}, which
considers a ``clapping mode'' oscillation between the two order
parameters. In our notation, as is clear from
Fig~\ref{fig_didmod}b, this mode is mainly the $\psi$ mode: we
claim that near a transition to the $d_{x^2-y^2} + i d_{xy}$
state, such a mode will be at high energy and overdamped; our
symmetry analysis also shows that it is decoupled from the proper
low energy mode, which is the $\phi$ oscillation. Further the field
independence of our normal mode frequencies does not agree with
\cite{bks}.
The main reason for these discrepancies seems the absence of
the $J({\bf k},\omega_n)$ term (\ref{e1},\ref{exprab})
in the Landau-Ginzburg free energy of Ref \cite{bks},
in their notation it would correspond to
$(\Delta_0^{\ast} \Delta_1 )^2 + h.c.$, which is allowed by
symmetry on the square lattice;
here $\Delta_{0,1}$ are the $d_{x^2-y^2}$ and $d_{xy}$ order
parameter components.

Finally, we discuss possible experimental signatures of the
$d+id$ instability.
A time-reversal symmetry breaking $d_{x^2-y^2} + i d_{xy}$
bulk ground state has not been detected so far;
so we assume that the system has $r>r_c$.
A low-energy $\phi$ oscillation mode should be well defined close to
the critical point, it is associated with chiral fluctuations
\cite{sigrist}.
A comparison with anomalies in transport measurements
(see Ref \cite{vzs2}) suggests an energy of order 20 K
for this mode.
Therefore we expect a corresponding signal in polarized
Raman \cite{raman} or neutron scattering \cite{neutron}
where the use of appropriate circular polarizations
allows a unique identification of chiral fluctuations.


\section{Conclusion}

In this paper, we have reviewed theories for quantum phase transitions
in $d$-wave superconductors which can cause strong damping
of low-energy quasiparticles.
All these theories are associated with a
quantum critical point between a $d$-wave superconductor and some
other superconducting state $X$ (Fig~\ref{fig_pd}b).
To explain the strong, inelastic scattering of nodal fermions
as observed in photoemission well below $T_c$, the state $X$ has to be
associated with a spin-singlet, zero momentum, fermion bilinear
order parameter.
We found that only two candidates for $X$ possessed a non-trivial
quantum critical point, a $d_{x^2-y^2}+is$ or a
$d_{x^2-y^2}+id_{xy}$ superconductor, where
the latter does not affect the antinodal fermions.
We have also considered the damping of antinodal fermions
arising from the proximity to a Neel ordered state.
The fermion scattering from antiferromagnetic fluctuations provides
an explanation for the broad lineshapes and the
``peak-dip-hump'' structure seen in experiment.

Finally we have discussed the collective modes of a
$d$-wave superconductor close to a zero temperature
$d_{x^2-y^2}+id_{xy}$ instability.
The only observable low-energy mode is associated with amplitude
fluctuations of the imaginary part of the $d_{xy}$ order parameter
component,
whereas the suggested ``clapping'' mode~\cite{bks} is found to be
at high energy and overdamped.


\section*{Acknowledgments}
We thank the US NSF (DMR 00--98226) and the
DFG (VO 794/1-1 and SFB 484) for support.

%

\end{document}